\documentclass[reprint,amsmath,amssymb, prl]{revtex4-2}

\usepackage[capitalise]{cleveref}
\usepackage{graphicx}
\usepackage{dcolumn}
\usepackage{bm}
\usepackage{xcolor}
\usepackage{verbatim} 
\usepackage{float}

\newcommand{\gv}[1]{\ensuremath{\mbox{\boldmath$ #1 $}}} 
\newcommand{\bv}[1]{\ensuremath{\boldsymbol{#1}}} 

\newcommand{\dev}{'}
\newcommand{\Ca}{\ensuremath{Cau}}
\newcommand{\order}[1]{\mathcal{O} \left( #1 \right)}

\begin{document}

\preprint{APS/123-QED}

\title{Soft streaming -- flow rectification via elastic boundaries}

\author{Yashraj Bhosale}
\affiliation{Mechanical Sciences and Engineering, University of Illinois at Urbana-Champaign, Urbana, IL 61801, USA}
\author{Tejaswin Parthasarathy}
\affiliation{Mechanical Sciences and Engineering, University of Illinois at Urbana-Champaign, Urbana, IL 61801, USA}
\author{Mattia Gazzola}
\email{mgazzola@illinois.edu}
\affiliation{Mechanical Sciences and Engineering, University of Illinois at Urbana-Champaign, Urbana, IL 61801, USA}

\date{\today}

\begin{abstract}
Viscous streaming is an efficient mechanism to exploit inertia at the microscale for flow control. While streaming from rigid features has been thoroughly investigated, when body compliance is involved, as in biological settings, little is known. Here, we investigate body elasticity effects on streaming in the minimal case of an immersed soft cylinder. Our study reveals an additional streaming process, available even in Stokes flows. Paving the way for advanced forms of flow manipulation, we illustrate how gained insights may translate to complex geometries beyond circular cylinders.
\end{abstract}

\maketitle



This paper examines the role of body elasticity in two-dimensional viscous streaming. Viscous streaming \cite{holtsmark1954boundary,lane1955acoustical,bertelsen1973nonlinear}, an inertial phenomenon, refers to the time-averaged, rectified steady flows that arise when an immersed body 
of length scale \(a\) undergoes small-amplitude oscillations in a viscous fluid. Long understood for rigid bodies of uniform curvature, such as cylinders~\cite{holtsmark1954boundary} or spheres~\cite{lane1955acoustical}, viscous streaming has found application in microfluidics~\cite{lutz2003microfluidics,lutz2005microscopic,marmottant2004bubble,lutz2006hydrodynamic,wang2011size,chong2013inertial,chen2014manipulation,klotsa2015propulsion,thameem2017fast,pommella2021enhancing}, from chemical mixing~\cite{liu2002bubble,lutz2003microfluidics,lutz2005microscopic,ahmed2009fast} to vesicle transport~\cite{marmottant2003controlled,marmottant2004bubble}, due to its ability to reconfigure flow and particle trajectories within short length --\(\order{100}\mu \)m-- and time --\(\order{10^{-3}}\)s-- scales~\cite{thameem2016particle,thameem2017fast}. 
Recent developments have then furthered opportunities in transport, separation or assembly, through the use of multi-curvature bodies and associated rich flow topologies~\cite{parthasarathy2019streaming,bhosale_parthasarathy_gazzola_2020,chan2021three,bhosale2021multi}.

Despite progress, no effort has so far systematically considered the role of body elasticity in viscous streaming. Yet, modulation by soft interfaces may be relevant in a multitude of settings, from pulsatile physiological flows~\cite{jalal2018three, jacob2021impact,parthasarathy2020simple} or conformal microfluidics~\cite{someya2016rise,heikenfeld2018wearable,bandodkar2019battery} to elastic mini-robots in fluids~\cite{park2016phototactic,ceylan2017mobile,huang2019adaptive,aydin2019neuromuscular}, with relevance to both medicine and engineering. Soft biological organisms, such as bacteria~\cite{spelman2017arbitrary} or larvae~\cite{gilpin2020multiscale}, may also take advantage of streaming for feeding or locomotion. Indeed, a back of the envelope calculation reveals that a millimeter-size aquatic organism beating its cilia at \(\sim \order{10}\) Hz would operate at the edge of viscous streaming viability. Supporting this hypothesis, steady flow patterns and velocities ($\sim 10^2-10^3~\mu\textrm{m}$/s) consistent with streaming have been observed in starfish and ribbon-worm larvae \cite{gilpin2020multiscale}, although being ascribed, perhaps inaccurately, to Stokes flow phenomena.

Motivated by these considerations, we dissect the effect of body elasticity on viscous streaming in the minimal setting of an immersed, oscillating hyperelastic circular cylinder.
The major outcome is that, in these conditions, the time-averaged streaming flow \( \langle \psi_1 \rangle\) reads
\begin{equation}
    \langle \psi_1 \rangle = \sin 2 \theta ~ \left[ \Theta(r) + \Lambda(r) \right]
\end{equation}
where \( r, \theta\) are cylindrical coordinates, $\Theta(r)$ is the classical rigid body solution from~\citet{holtsmark1954boundary}, and $\Lambda(r)$ is a novel, independent contribution from body elasticity. 

%
%
\begin{figure}[t]
\includegraphics[width=0.9\linewidth]{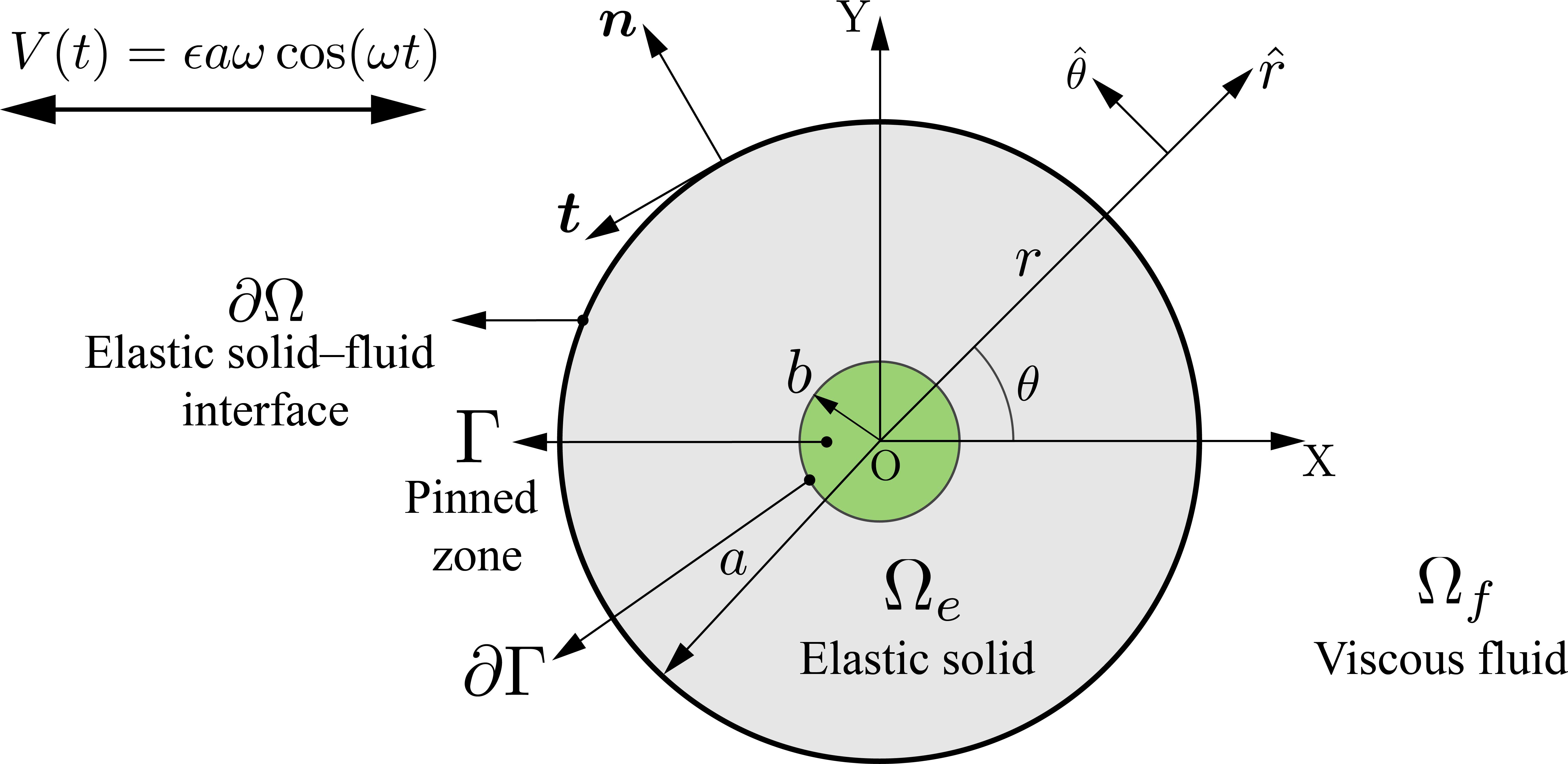}
\caption{\label{fig:setup} Problem setup. Elastic solid cylinder $\Omega_e$ of radius \(a\) with a rigid inclusion (pinned zone $\Gamma$ of radius $b$), immersed in the viscous fluid $\Omega_f$. The cylinder is
exposed to an oscillatory flow with far-field velocity $V(t) = \epsilon a \omega \cos (\omega t)$.
}
\end{figure}
%
%
\par
The above result is obtained by considering the setup shown in~\cref{fig:setup}, where a 2D visco-elastic solid 
cylinder $\Omega_e$ with radius \( a \) is immersed in a viscous fluid $\Omega_f$. The fluid oscillates 
with velocity $V(t) = \epsilon a \omega \cos \omega t$, where $\epsilon$, $\omega$ and $t$ represent the non-dimensional 
amplitude, angular frequency and time, respectively. We `pin' the cylinder's centre using a rigid inclusion $\Gamma$ of radius $b < a$, to kinematically enforce zero strain and velocity near the cylinder's centre.
We denote by \(\partial\Omega\) and \( \partial\Gamma \) the boundary between the elastic solid and viscous fluid, 
and the boundary of the pinned zone, respectively.
\par
In this setup, we assume fluid and solid to be isotropic, incompressible and of 
constant density. Furthermore, we assume the fluid to be Newtonian, with kinematic viscosity $\nu_f$
and density $\rho_f$.
We assume that the solid exhibits visco-hyperelastic neo-Hookean behaviour, characteristic of soft biological materials~\cite{bower2009applied}, with shear modulus $G$,  kinematic viscosity $\nu_e$ and density $\rho_e$.

The dynamics in the elastic and fluid phases, 
are described by the incompressible Cauchy momentum equations, non-dimensionalized using the characteristic scales of velocity $V = \epsilon a \omega$, length $L = a$  and time $T = 1 / \omega$ 
\begin{equation}
    \begin{aligned}
        \label{eqn:gov_eqns_nondim}
        \textrm{Incomp.}
        &\begin{cases}
         \gv{\nabla} \cdot \gv{v} = 0, ~\gv{x}\in \Omega_f \cup \Omega_e 
        \end{cases}\\
        \textrm{Fluid}
        &\begin{cases}        
        \frac{\partial {\gv{v}}}{\partial {t}} + \epsilon 
        ({\gv{v}} \cdot {\bv{\nabla}}) {\gv{v}} = \frac{1}{M^2} 
        \left(-{\nabla}{p} + {\nabla^2} {\gv{v}} \right),~{\gv{x}}\in\Omega_f
        \end{cases}\\
        \textrm{Solid}
        &\begin{cases} 
        \alpha \Ca \left( \frac{\partial {\gv{v}}}{\partial {t}} + \epsilon 
        ({\gv{v}} \cdot {\bv{\nabla}}) {\gv{v}} \right) = \frac{\Ca}{M^2}
        \left( -{\nabla} {p} + \beta {\nabla^2} {\gv{v}} \right) \\ \hspace{3.5cm}+
        {\bv{\nabla}} \cdot ({\bv{F}} {\bv{F}}^T)\dev 
        ,~{\gv{x}}\in\Omega_e,
        \end{cases}\\
    \end{aligned}
\end{equation}
%
%
%
where $\gv{v}$ and $p$ are the velocity and pressure fields, and \(\bv{F}\) is the deformation gradient tensor, 
defined as $\bv{F} = \bv{I} + \bv{\nabla} \gv{u}$, where \(\bv{I}\) is the identity, $\gv{u} = \gv{x} - \gv{X}$ is the material 
displacement field, and $\gv{x}$, $\gv{X}$ are the position of a material 
point after deformation and at rest, respectively. The prime symbol \(\dev\) on a tensor denotes its deviatoric. In addition, the following non-dimensional groups naturally appear: scaled oscillation amplitude \(\epsilon \), Womersley number $M = a \sqrt{\rho_f \omega / \mu_f}$, Cauchy number $\Ca = \epsilon \rho_f a^2 \omega^2 / G$, 
density ratio \(\alpha = \rho_e / \rho_f\) and viscosity ratio \( \beta = \mu_e / \mu_f\). Physically, $M$ represents 
the ratio of inertial to viscous forces, and \(\Ca\) represents the ratio of inertial to elastic forces. Thus, increasing $M$ indicates an inertia-dominated environment, and increasing $\Ca$ implies a softer body.
%
%
The equations are then closed using the boundary conditions
\begin{align*}
    \textrm{Pinned zone}
        &\begin{cases}
            \tag{3.1}
            \label{eqn:bcs_pinned}
            \gv{u} = 0, \gv{v} = 0,~\gv{x}\in \Gamma
        \end{cases}\\
    \tag{3.2}
    \label{eqn:bcs_velocity}
    \textrm{Interface velocity}
        &\begin{cases}
            \gv{v}_e = \gv{v}_f,~\gv{x}\in\partial\Omega        
        \end{cases}\\
    \tag{3.3}
    \label{eqn:bcs_stresses}
    \textrm{Interface stresses}
        &\begin{cases}
            \bv{\sigma}_{f} = -p\bv{I} +  (\bv{\nabla} \gv{v} + \bv{\nabla} \gv{v}^T), ~\gv{x}\in\Omega_f \\
            \bv{\sigma}_{e} = -p\bv{I} + \beta (\bv{\nabla} \gv{v} + \bv{\nabla} \gv{v}^T)
                \\ ~~~~+ \frac{M^2}{\Ca}(\bv{F} \bv{F}^T)\dev,
            ~\gv{x}\in\Omega_e, \\
            \gv{n} \cdot \bv{\sigma}_e \cdot \gv{n} = 
            \gv{n} \cdot \bv{\sigma}_f \cdot \gv{n},~\gv{x}\in\partial\Omega\\
            \gv{n} \cdot \bv{\sigma}_e \cdot \gv{t} = 
            \gv{n} \cdot \bv{\sigma}_f \cdot \gv{t},~\gv{x}\in\partial\Omega\\
        \end{cases}\\
    \tag{3.4}
    \label{eqn:bcs_farfield}
    \textrm{Far-field}
        &\begin{cases}
            \gv{v}(|\gv{x}| \to \infty) =  \cos \omega t ~\hat{i},~\gv{x}\in \Omega_{f},
        \end{cases}
\end{align*}
where \cref{eqn:bcs_pinned} is the rigid pin constraint,~\cref{eqn:bcs_velocity} is the no-slip condition,~\cref{eqn:bcs_stresses} enforces continuity of stresses, and~\cref{eqn:bcs_farfield} is the far-field flow. 
Next, we identify relevant parameter ranges and solve~\cref{eqn:gov_eqns_nondim} accordingly, using perturbation theory.

Typically, in viscous streaming applications, the scaled oscillation amplitude is $\epsilon \ll {1}$~\cite{holtsmark1954boundary,bertelsen1973nonlinear,lutz2005microscopic}, and the Womersley number is $M \geq \order{1}$ \cite{marmottant2004bubble,lutz2006hydrodynamic}. Additionally, density $\alpha $ and viscosity $\beta$ ratios are $\sim \order{1}$. The Cauchy number $\Ca$ requires careful consideration. For a rigid body $\Ca = 0$, while for an elastic body $\Ca > 0$, with $\Ca \ll 1$ implying weak elasticity.  From a mathematical perspective, dealing with $\Ca \geq \order{1}$ is challenging due to the highly non-linear nature of hyperelastic materials. Here, we assume that the cylinder is only weakly elastic, and in particular that $\Ca = \kappa \epsilon$, where $\kappa = \order{1}$. This assumption simplifies the asymptotic treatment, slaving \( \Ca \) to \( \epsilon\) so that both are equally small and tend to zero at the same rate.


We then look for asymptotic solutions of~Eqs.(\ref{eqn:gov_eqns_nondim}) by perturbing all relevant fields as series of powers of $\epsilon$. We derive the leading order solution $\order{1}$, which reduces to a rigid cylinder in a purely oscillatory flow governed by the unsteady Stokes equation~\cite{holtsmark1954boundary}. The next order solution $\order{\epsilon}$ is derived in two steps. First, we obtain the deformation of the elastic solid due to the leading order flow. Second, we use this deformation to
determine the boundary conditions for the flow at $\order{\epsilon}$, thus incorporating elasticity effects into the streaming solution. Steps are mathematically outlined below, with details in the SI.


We start by perturbing to \( \order{\epsilon}\) all physical quantities \( q\), which include \(\gv{v}\), \(\gv{u}\), \(p\),  \( \Omega \), \(\gv{n}\),  \(\gv{t}\), as
\begin{equation}
    \label{eqn:perturb}
    q \sim q_0 + \epsilon q_1 + \order{\epsilon^2}
\end{equation}
and substitute them in Eqs.~\ref{eqn:gov_eqns_nondim}. Subscripts (0, 1, ...) indicate the solution order. Then, we adopt the more convenient cylindrical coordinate system $(r, \theta)$, with radial coordinate $r$, angular coordinate $\theta$, and origin at the center of the cylinder. Horizontal axis direction $\gv{i}$ corresponds to $\theta = 0$. 
%
At leading order \( \order{1}\), the governing equations and boundary conditions in the solid reduce to
\begin{equation}
        \bv{\nabla} \cdot ((\bv{I} + \bv{\nabla} \bv{u}_0) (\bv{I} + \bv{\nabla} \bv{u}_0)^T)\dev = 0,
        ~~ r \leq 1;~~~
        \gv{u}_0|_{r = \zeta} = 0
    \label{eqn:gov_eqns_nondim_solid_psi0}
\end{equation}
where $\zeta = b / a$ is the non-dimensional radius of the pinned zone.
Since at this order $\Ca = \kappa \epsilon = 0$, the solution 
of Eqs.~\ref{eqn:gov_eqns_nondim_solid_psi0} is the fixed, rigid body cylinder
\begin{equation}
    \partial\Omega_0 = r = 1;~~
    \gv{u}_0 = 0,~~ 
    \gv{v}_{0} = \frac{\partial \gv{u}_0}{dt} = 0 ,~~ r \leq 1.
    \label{eqn:solid_zero}
\end{equation}
With these leading order boundary conditions the governing equations in the fluid reduce to
\begin{equation}
    \begin{aligned}
        M^2 \frac{\partial \nabla^2 \psi_0}{\partial t} &= \nabla^4 \psi_{0}~~~~ r \geq 1 \\
        v_{0,r}|_{r = 1} = \frac{1}{r} \frac{\partial \psi_0}{\partial \theta}\biggr|_{r = 1} = 0 &;~
        v_{0, \theta}|_{r = 1} = -\frac{\partial \psi_0}{\partial r}\biggr|_{r = 1} = 0 \\
        v_{0, r}|_{r \to \infty} = \cos \theta ~ \cos t &;~
        v_{0, \theta}|_{r \to \infty} = -\sin \theta ~ \cos t,
    \end{aligned}
    \label{eqn:gov_eqns_nondim_fluid_psi0}
\end{equation}
where \(\psi\) is the streamfunction defined as $\gv{v} = \bv{\nabla} \times \psi$. This system (Eq.~\ref{eqn:solid_zero}, \ref{eqn:gov_eqns_nondim_fluid_psi0}) is a rigid cylinder immersed in an oscillating unsteady Stokes flow, which has the exact analytical solution~\cite{holtsmark1954boundary}
\begin{equation}
    \label{eqn:soln_psi0}
    \psi_0 = \frac{\sin\theta}{2} \left( r + \frac{H_2(m)}{r H_0(m)} - 
    \frac{2H_1(m r)}{m H_0(m)}\right) e^{-it} + c.c. ,~r \geq 1
\end{equation}
where $i = \sqrt{-1}$, and $m = \sqrt{i} M$. Here, $H_i$ and $c.c.$ refer to the $i^{\textrm{th}}$ order Hankel function of first kind and complex conjugate. The leading order field $\psi_0$ in the fluid is purely oscillatory, thus no steady streaming is observed at $\order{1}$, as expected \cite{holtsmark1954boundary, bertelsen1973nonlinear}. Additionally, no effects of elasticity manifest on the flow at this order.

We then proceed to the next order of approximation $\order{\epsilon}$, where we instead do expect steady streaming to emerge and elasticity to play a role. At $\order{\epsilon}$, the solid governing equations reduce (SI, Eqs.~42, 49) to
\begin{equation}
    \label{eqn:gov_eqns_nondim_solid1_simp_psi}
    \nabla^4 \psi_{e,1} = 0,~~~~\gv{x}\in\Omega_{e}
\end{equation}
where we have defined the strain function $\psi_{e}$, so that $\gv{u} = \nabla \times \psi_{e}$ (similar to the streamfunction $\psi$).~\Cref{eqn:gov_eqns_nondim_solid1_simp_psi} shows how the specific choice of solid elasticity model is irrelevant at \(\order{\epsilon}\), since all non-linear stress-strain responses drop out due to linearization. \Cref{eqn:gov_eqns_nondim_solid1_simp_psi} is further complemented by the boundary conditions at the pinned zone interface
\begin{equation}
    \label{eqn:pinned1_cyl_psi}
    u_{1, r} = \left. \frac{1}{r}\frac{\partial \psi_{e,1}}{\partial \theta}\right|_{r = \zeta} = 0;~~~~
    u_{1, \theta} = \left.-\frac{\partial \psi_{e,1}}{\partial r}\right|_{r = \zeta} = 0.
\end{equation}
Now, the flow solution at \(\order{1}\) exerts interfacial stresses on the solid, which at \( \order{\epsilon}\), is no longer rigid but instead deforms. This process is driven by~\cref{eqn:bcs_stresses}, which yields the following radial and tangential stress conditions 
\begin{equation}
    \begin{aligned}
        \label{eqn:elastic_bcs_nondim_solid1_cyl_psi}
            \frac{M^2}{\kappa} \left.\frac{\partial}{\partial r}\left( \frac{1}{r}\frac{\partial\psi_{e,1}}{\partial \theta}\right)\right|_{r = 1} = 
            \left.\frac{\partial v_{0,r}}{\partial r}\right|_{r = 1} \\
            \frac{M^2}{\kappa} \left. \left( \frac{1}{r^2}\frac{\partial^2 \psi_{e,1}}{\partial \theta^2} - r \frac{\partial}{\partial r}\left( \frac{1}{r}\frac{\partial \psi_{e,1}}{\partial r}\right)
                     \right) \right|_{r = 1} = \\
            \left. \left( \frac{1}{r}\frac{\partial v_{0,r}}{\partial \theta} + \frac{\partial
                v_{0,\theta}}{\partial r} - \frac{v_{0,\theta}}{r} \right) \right|_{r = 1},
    \end{aligned}
\end{equation}
where the LHS corresponds to the solid phase (Eq.~\ref{eqn:gov_eqns_nondim_solid1_simp_psi}) and the RHS to the fluid phase (Eq.~\ref{eqn:soln_psi0}), both evaluated at the leading order interface \( r = 1\). 
We note that although the solid interface does deform, the use of \( r = 1\) is not inconsistent. Indeed, as shown in the SI (Eq.~44--46), this approximation induces higher order $\order{\epsilon^2}$ errors in the boundary stresses evaluation.
The flow quantities on the RHS can be then directly evaluated
\begin{equation}
    \begin{aligned}
        \left.\frac{\partial v_{0,r}}{\partial r}\right|_{r = 1} &= 0 \\
            \left.\left( \frac{1}{r}\frac{\partial v_{0,r}}{\partial \theta} + \frac{\partial
                v_{0,\theta}}{\partial r} - \frac{v_{0,\theta}}{r} \right)\right|_{r = 1} &= \sin{\theta} ~ F(m) ~ e^{-it} + c.c.    
    \end{aligned}
    \label{eqn:zero_stress}
\end{equation}
with 
\begin{equation} 
    \label{eqn:f}
    F(m) = {-m H_{1}(m)}/{H_0(m)}.
\end{equation}
Once \cref{eqn:zero_stress,eqn:f} are substituted back in the boundary conditions of \cref{eqn:elastic_bcs_nondim_solid1_cyl_psi}, the biharmonic~\cref{eqn:gov_eqns_nondim_solid1_simp_psi} can be solved to obtain the \( \order{\epsilon}\) solid displacement field
\begin{equation}
    \begin{aligned}
    \label{eqn:strain_func}
    \psi_{e,1} &= \frac{\kappa}{M^2} \sin{\theta} (r) \left( c_1 + \frac{c_2}{r^2} +  c_3 r^2 + c_4 \ln(r) \right) F(m) ~ e^{-it} \\ 
    & + c.c.,
    \end{aligned}
\end{equation}
where the expressions for \(c_1, c_2, c_3, c_4 \) are reported in the SI.
\Cref{eqn:strain_func} represents the \(\order{\epsilon}\) solid displacement field both in the bulk $\Omega_e$ and at the boundary $\partial \Omega$, which directly affects the \(\order{\epsilon}\) flow. At \(\order{\epsilon}\), the flow governing equation, in streamfunction form, reads \cite{holtsmark1954boundary}
\begin{equation}
\label{eqn:gov_eqns_nondim_fluid_psi1}
M^2 \frac{\partial \nabla^2 \psi_1}{\partial t} + 
M^2 \left( \left( \gv{v}_0 \cdot \bv{\nabla} \right) \nabla^2 \psi_0 \right)
= \nabla^4 \psi_{1},~~~~ r \geq 1.
\end{equation}
Since we are interested in steady streaming, we consider the time average \begin{equation}
    \begin{aligned}
        \label{eqn:gov_eqns_nondim_fluid_psi1_steady}
        \nabla^4 \langle \psi_{1} \rangle &= 
        M^2
        \underbrace{\langle \left( \gv{v}_0 \cdot \bv{\nabla} \right) \nabla^2 \langle \psi_0 \rangle \rangle}_{\textrm{RHS}}
        ,~~~~ r \geq 1, \\
    \end{aligned}
\end{equation}
where the RHS can be rewritten using~\cref{eqn:soln_psi0} to yield
\begin{equation}
    \begin{aligned}
        \nabla^4 \langle \psi_{1} \rangle = 
        \sin 2 \theta ~ \rho(r)
        ,~~~~ r \geq 1, \\
        \rho(r) = -\frac{M^4}{2} ~ \textrm{Im}\left[ \frac{H_2(m r)}{H_0(m)} + \frac{H_2(m) H_0^*(m
        r)}{H_0^2(m) r^2} \right. \\
        \left. + 2 \frac{H_0(m r) H_2^*(m r)}{H_0^2(m)}\right],
    \end{aligned}
    \label{eqn:rey_stresss}
\end{equation}
with $\textrm{Im}[\cdot]$ representing the imaginary part. To solve this equation, we first recall the far-field boundary conditions
\begin{equation}
    \begin{aligned}
        \label{eqn:fffar}
        \left.\frac{1}{r} \frac{\partial \langle \psi_1 \rangle}{\partial \theta}\right|_{r \to \infty} =
        \left.\frac{\partial \langle \psi_1 \rangle}{\partial r}\right|_{r \to \infty} = 0.
    \end{aligned}
\end{equation}
Next, we recall the no-slip boundary condition of \cref{eqn:bcs_velocity} that needs to be enforced at the $\order{\epsilon}$ accurate solid--fluid interface (SI, Eq.~62)
\begin{equation}
    \begin{aligned}
    \label{eqn:fluid_bc1_pre}
    \left. \gv{v}_{e} \right|_{\partial \Omega} &= 
    \left. \gv{v}_{e} \right|_{r = 1 + \epsilon u_{1, r}} + \order{\epsilon^2}\\  &= 
    \left. \gv{v}_{f} \right|_{\partial \Omega}
    =\left. \gv{v}_{f} \right|_{r = 1 + \epsilon u_{1, r}} + \order{\epsilon^2},
    \end{aligned}
\end{equation}
\begin{figure*}[!ht]
\includegraphics[width=\linewidth]{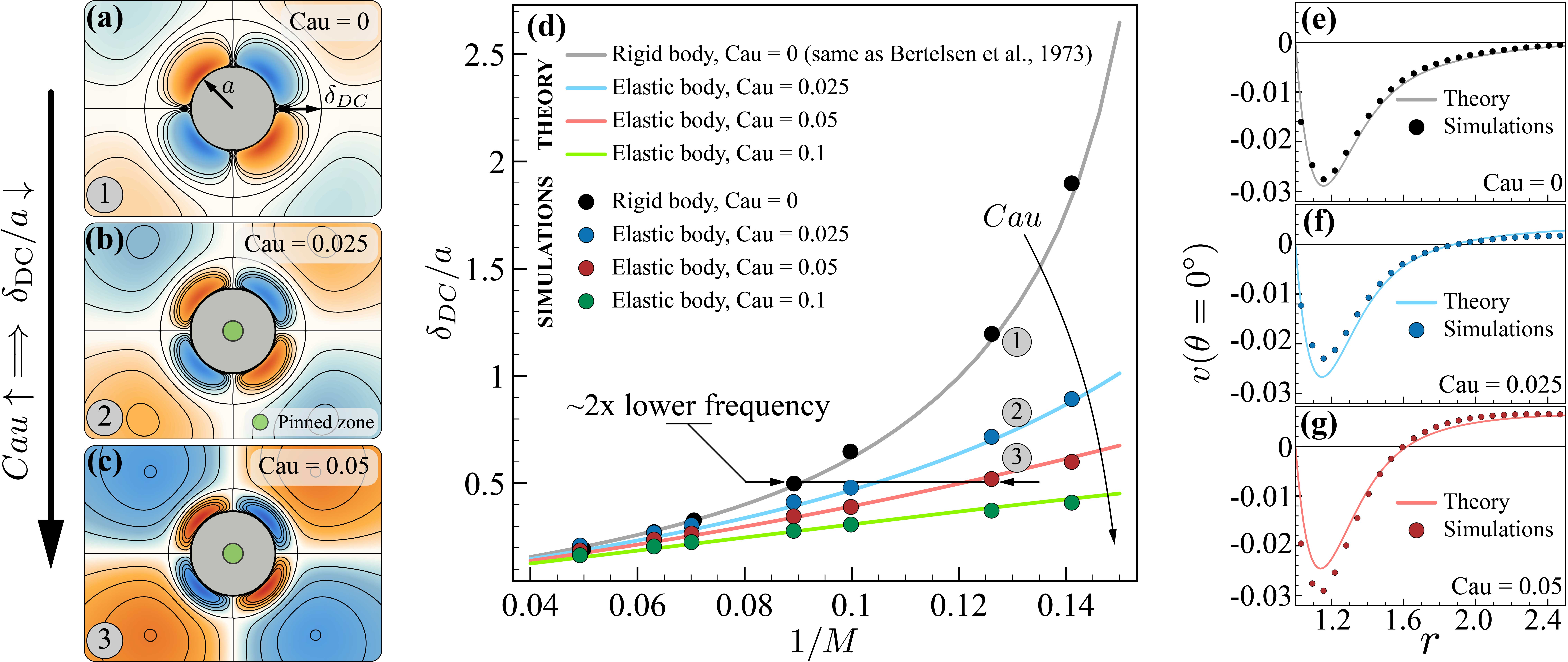}
\caption{\label{fig:DClayer} Effect of elasticity on streaming flow. Time averaged streamline patterns (blue/orange represent clockwise/anti-clockwise rotating regions) depicting streaming response at $M \approx 8$ with increasing softness $\Ca$: (a) rigid limit ($\Ca = 0$), (b) $\Ca = 0.025$ and (c) $\Ca = 0.05$. Non-dimensional radius of the pinned zone (green cylinder) is set at $\zeta = 0.2$ throughout the study. For effects of $\zeta$ variation on streaming topology, refer to SI. (d) Normalized DC layer thickness $\delta_{DC} / a$ vs. inverse of Womersley number ($1 / M$) from theory and simulations, for varying body elasticity $\Ca$. Radial decay of velocity magnitude along $\theta = 0^{\circ}$ from theory and simulations, with increasing softness $\Ca$: (e) rigid limit ($\Ca = 0$), (f) $\Ca = 0.025$ and (g) $\Ca = 0.05$. For simulation details, refer to SI.
}
\end{figure*}

\noindent where we highlight how, at $\order{\epsilon}$, the cylinder interface is no longer fixed at $r = 1$, but deforms as $r' = 1 + \epsilon u_{1, r}$. Here, $u_{1, r}$ is the $\order{\epsilon}$ accurate deformation field computed by injecting \cref{eqn:strain_func} into $\gv{u}_1 = \nabla \times \psi_{e,1}$. To enforce \cref{eqn:fluid_bc1_pre}, while maintaining an analytically tractable formulation, we replace the boundary flow velocity $\left. \gv{v}_f \right|_{r = r'}$ on the temporally moving interface $r'$ with \textit{the velocity that the flow would need to see on the fixed interface $r = 1$ to respond equivalently}. We achieve this by Taylor expanding $\left. \gv{v}_{f} \right|_{r = r'}$ about $r = 1$ (SI, Eq. 64--66)
%
\begin{equation}
    \begin{aligned}
\label{eqn:fluid_bc1_expand}
\left. \gv{v}_{f} \right|_{r = 1 + \epsilon u_{1,r}} &= \left. \left( \epsilon \gv{v}_{f,1} + 
    \epsilon \frac{\partial \gv{v}_{f,0}}{\partial r} u_{1,r} \right) \right|_{r = 1} +
        \order{\epsilon^2}.
    \end{aligned}
\end{equation}
Similarly, $\left. \gv{v}_{e} \right|_{r = 1 + \epsilon u_{1, r}}$ (LHS of Eq.~\ref{eqn:fluid_bc1_pre}) can be computed to $\order{\epsilon}$ accuracy as $\left. \partial \gv{u}_{1,r} / \partial t \right|_{r = 1}$ (SI, Eq. 63). 
Given that $u_{1,r}$ (Eq.~\ref{eqn:strain_func}) and $\partial v_{f,0} / \partial r$ (Eq.~\ref{eqn:soln_psi0}) are known, we can plug \cref{eqn:fluid_bc1_expand} into \cref{eqn:fluid_bc1_pre} to obtain $v_{f,1}$ (SI, Eq.~62--67). Time averaging yields
\begin{equation}
    \begin{aligned}
        \label{eqn:bc_nondim_fluid_psi1_steady}
        \left. \langle v_{1, r} \rangle \right|_{r = 1} = 
        \left. \frac{1}{r} \frac{\partial \langle \psi_1 \rangle}{\partial \theta}\right|_{r = 1} &= 0~\\
        \left. -\langle v_{1, \theta} \rangle \right|_{r = 1} = \left.\frac{\partial \langle \psi_1 \rangle}{\partial r}\right|_{r = 1} &= \frac{\kappa}{M^2} \sin 2 \theta ~ G_1(\zeta) F(m) F^*(m)
    \end{aligned}
\end{equation}
with
\begin{equation} 
    \label{eqn:g}
    G_1(\zeta) = 0.5 \left( \frac{(\zeta^2 + 1) ln (\zeta)}{\zeta^2 - 1} - 1\right).
\end{equation}
\Cref{eqn:bc_nondim_fluid_psi1_steady} tells us that, from the fluid perspective, the no-slip condition on the moving interface $r'$ can be equivalently seen as a rectified tangential slip velocity ($\langle v_{1, \theta} \rangle|_{r = 1} \neq 0$) on the leading order, fixed interface $r = 1$. 
In our case, this slip velocity stems from solid elasticity and modifies the Reynolds stresses---$\sin 2 \theta ~ \rho(r)$--- associated with the rigid body (Eq.~\ref{eqn:rey_stresss}), thus altering the overall streaming flow response. We remark that this slip is independent of the Navier--Stokes non-linear inertial advection. Hence, streaming can be generated even in the Stokes limit, unlike for rigid bodies. 

Given the steady flow of Eq.~(\ref{eqn:rey_stresss})
and boundary conditions of \cref{eqn:fffar,eqn:bc_nondim_fluid_psi1_steady}, the streaming solution can finally be written as
\begin{equation}
    \label{eqn:psi_final_form}
    \langle \psi_1 \rangle = \sin 2 \theta ~ \left[ \Theta(r) + \Lambda(r) \right].
\end{equation}
Here, $\Theta(r)$ is the classical rigid body contribution from~\citet{holtsmark1954boundary}
\begin{equation}
    \begin{aligned}
    \label{eqn:psi_rigid}
        &\Theta(r) = -\frac{r^{4}}{48} \int_{r}^{\infty} \frac{\rho(\tau)}{\tau} ~ \mathrm{d} \tau + \frac{r^{2}}{16} \int_{r}^{\infty} \tau \rho(\tau) ~ \mathrm{d} \tau \\
        &+ \frac{1}{16}\left(\int_{1}^{r} \tau^{3} \rho(\tau) ~ \mathrm{d} \tau + \int_{1}^{\infty} \frac{\rho(\tau)}{\tau} ~ \mathrm{d} \tau -2 \int_{1}^{\infty} \tau \rho(\tau) ~ \mathrm{d} \tau\right) \\
        &+ \frac{1}{r^{2}}\left(-\frac{1}{48} \int_{1}^{r} \tau^{5} \rho(\tau) ~ \mathrm{d} \tau - \frac{1}{24} \int_{1}^{\infty} \frac{\rho(\tau)}{\tau} ~ \mathrm{d} \tau \right. \\
        &\left. + \frac{1}{16} \int_{1}^{\infty} \tau \rho(\tau) ~ \mathrm{d} \tau\right)
    \end{aligned}
\end{equation}
and $\Lambda(r)$ is the new elastic modification
\begin{equation}
    \label{eqn:psi_elastic}
    \Lambda(r) = 0.5 \frac{\kappa}{M^2} ~ G_1(\zeta) F(m) F^*(m) \left( 1 - \frac{1}{r^2} \right)
\end{equation}
with $G_1(\zeta)$ and $F(m)$ given in~\cref{eqn:g} and  \cref{eqn:f}.
This concludes our theoretical analysis.
\begin{figure*}[ht]
\includegraphics[width=1.0\linewidth]{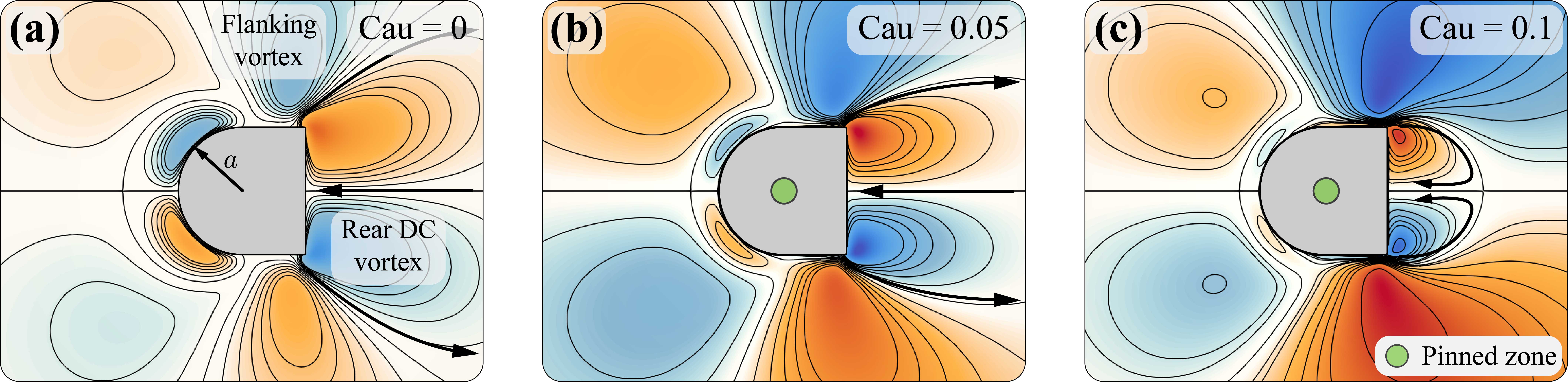}
\caption{\label{fig:bullet} Extension of compliance-induced streaming to generalised, multiple curvature bodies.
Time averaged simulation flow topologies for a bullet---formed by hybridizing a circle of radius $a$ with a square of side length $2a$---at $M \approx 8$, with varying body elasticity $\Ca$: (a) rigid limit $\Ca = 0$, (b) $\Ca = 0.05$ and (c) $\Ca = 0.1$. Increasing body softness results in contraction and strengthening of the rear DC vortex on the square side, consistent with our theoretical insights.}
\end{figure*}
\par
Next, we compare our theory against direct numerical simulations \cite{bhosale2021remeshed} and known analytical results in the rigidity limit~\cite{holtsmark1954boundary,bertelsen1973nonlinear}. For a rigid cylinder (\( \Ca = 0\)) oscillating at $M \approx 8$, numerical time-averaged streamlines are shown in~\cref{fig:DClayer}(a). We highlight the four-fold symmetry and the presence of a well-defined direct circulation (DC) layer of thickness $\delta_{DC}$. \citet{holtsmark1954boundary} predicts this flow topology, as well as an increase of $\delta_{DC}$ with $1 / M$ until divergence, at which point the DC layer extends to infinity. This behaviour is recovered by our theory when \( \Ca = 0\) (i.e. \( \Lambda = 0\)), and by simulations (black line/dots in~\cref{fig:DClayer}d). As the cylinder becomes soft (\( \Ca > 0\)), four-fold symmetry is preserved ($\sin 2 \theta$ in Eq.~\ref{eqn:psi_final_form}), but $\delta_{DC}$ contracts on account of the elastic term $\Lambda \neq 0$. This is confirmed by simulations across a range of $\Ca$, as seen in~\cref{fig:DClayer}(b-d). Thus, an elastic body can access streaming flow configurations equivalent to rigid objects exposed to significantly higher oscillation frequencies (e.g., $\sim 2 \times$ at \( \Ca = 0.05\),~\cref{fig:DClayer}d). We conclude our validation by reporting in \cref{fig:DClayer}(e--g) theoretical and simulated radially-varying, time-averaged velocities \( |\langle v \rangle| \) at \(\theta = 0^{\circ} \), noting close agreement.
\par
Finally, we demonstrate how gained theoretical intuition extends to geometries of multiple curvatures. We consider the shape of \cref{fig:bullet}, previously designed \cite{bhosale_parthasarathy_gazzola_2020} to attain streaming flows favorable to particle transport \cite{parthasarathy2019streaming} and separation \cite{bhosale2021multi}. Both applications rely on the presence of flanking and rear vortices, and performance is improved by strengthening the vortices via increasing oscillation frequencies \cite{bhosale_parthasarathy_gazzola_2020}. \Cref{fig:bullet} shows how the same process can alternatively be achieved by increasing softness only. As a result, the same flow topologies of \cite{bhosale_parthasarathy_gazzola_2020} are obtained in \cref{fig:bullet} for frequencies $\sim 4 \times$ lower.

In summary, we derived a viscous streaming theory for the case of an elastic cylinder, and validated it computationally. Our study reveals an additional, tunable mode of streaming, accessible through material compliance and available even in Stokes flow. We demonstrate its use for flow control in the case of a previously designed streaming body of multiple curvatures, to illustrate application potential in microfluidics or microrobotics, in conjunction with the use of elastomeric or biological materials. Further, the fact that compliance enables streaming effects at frequencies significantly lower than rigid bodies, supports the hypothesis that biological creatures, speculated to operate at the edge of viscous streaming viability, may instead take full advantage of it thanks to their softness.



\bibliography{cfs_lit}

\end{document}